\documentclass[aps,prd,groupedaddress,showpacs,floatfix,12pt]{revtex4}
\usepackage{graphicx}
\usepackage{dcolumn}
\usepackage{bm}
\begin{document}
\title{Replay to hep-ph/0606266}
\author{N.~N.~Achasov}
\email{achasov@math.nsc.ru}
\affiliation{Laboratory of  Theoretical Physics,\\
S.~L.~Sobolev
Institute for Mathematics,\\
Academician Koptiug Prospekt, 4\\
 Novosibirsk, 630090, Russia}

\date{\today}

\begin{abstract}
 The $\phi\to\gamma\pi\pi$ amplitude  is found from the
$e^+e^-\to\gamma\pi\pi$ amplitude suggested in  Ref.
\cite{anisovich}. It is shown that the found amplitude differs
essentially from that studied in Refs.
\cite{anisovich,anisovich05}. It is noticed that the suggestion of
the author  of Ref. \cite{anisovich} about the phase of the
$\phi\to\gamma\pi\pi$ amplitude is misleading.
\end{abstract}
\vspace*{1cm}
\maketitle \vspace*{1cm}

First, I thank  V.V. Anisovich for his attention \cite{anisovich}
to my paper \cite{achasov}.

Let us rewrite Eq. (4) from Ref.  \cite{anisovich} in the real
case, that is, taking into account the widths of the resonances
\begin{eqnarray}
\label{four} &&
A^{(e^+e^-\to\gamma\pi\pi)}_{\mu\alpha}(s_V\,,\,s_S\,,\,0)=\left
(g_{\mu\alpha} - \frac{2q_\mu P_{V\alpha}}{s_V-s_S}\right
)\nonumber\\[9pt] &&\times\Biggl\{
\frac{G_{e^+e^-\to\phi}}{s_V-M_\phi^2 +
\imath\sqrt{s_V}\Gamma_\phi (s_V)}\Biggl [\frac{A_{\phi\to\gamma
f_0}\left (m_\phi^2\,,\,m_{f_0}^2\,,0\right )\,g_{f_0\to\pi\pi}
}{s_S-M_0^2 + \imath g^2_\pi\rho_{\pi\pi}(s_S)+ \imath
g^2_K\rho_{K\bar K}(s_S)}+ B_\phi\left
(m_\phi^2\,,\,s_S\,,\,0\right )\Biggr ]\nonumber\\[9pt]
&&  +
\frac{B_{f_0}\left(s_V\,,\,m_{f_0}^2\,,\,0\right)\,g_{f_0\to\pi\pi}}{s_S-M_0^2+
\imath g^2_\pi\rho_{\pi\pi}(s_S)+ \imath g^2_K\rho_{K\bar K}(s_S)
}+ B_0\left (s_V\,,\,s_S\right)\Biggr\}
\end{eqnarray}
where $\mu$ and $\alpha$ refer to the initial vector state (total
momentum $P_V$ and $P_V^2=s_V$) and photon (momentum $q$ and
$q^2=0$),  $\left (P_V-q\right )^2=s_S$ and $\left (P_Vq\right
)=\left (s_V-s_S\right )/2$\ .

 Then the regularity  condition of
$A^{(e^+e^-\to\gamma\pi\pi)}_{\mu\alpha}(s_V\,,\,s_S\,,\,0)$  at
$s_V=s_S$  has the form

\begin{eqnarray}
\label{five}&&  \frac{G_{e^+e^-\to\phi}}{s_V-M_\phi^2 + \imath
(s_V)\Gamma_\phi\sqrt{s_V }}\Biggl [\frac{A_{\phi\to\gamma
f_0}\left (m_\phi^2\,,\,m_{f_0}^2\,,0\right )\,g_{f_0\to\pi\pi}
}{s_V-M_0^2 + \imath g^2_\pi\rho_{\pi\pi}(s_V)+ \imath
g^2_K\rho_{K\bar K}(s_V)} + B_\phi\left
(m_\phi^2\,,\,s_V\,,\,0\right )\Biggr ] \nonumber\\[9pt]
 &&
  +
\frac{B_{f_0}\left(s_V\,,\,m_{f_0}^2\,,\,0\right)\,g_{f_0\to\pi\pi}}{s_V-M_0^2+
\imath g^2_\pi\rho_{\pi\pi}(s_V)+ \imath g^2_K\rho_{K\bar K}(s_V)
}+ B_0\left (s_V\,,\,s_V\right)=0\,.
\end{eqnarray}

When confined only to the $\phi$ meson contribution in Eqs.
(\ref{four})  and  Eq. (\ref{five}), one can find only the trivial
solution of Eq. (\ref{five})
\begin{equation}
\label{phi} \frac{A_{\phi\to\gamma f_0}\left
(m_\phi^2\,,\,m_{f_0}^2\,,0\right )\,g_{f_0\to\pi\pi}}{s_V-M_0^2 +
\imath g^2_\pi\rho_{\pi\pi}(s_V)+ \imath g^2_K\rho_{K\bar K}(s_V)}
 + B_\phi\left
(m_\phi^2\,,\,s_V\,,\,0\right )=0,
\end{equation}
which implies that there is not the $\phi$ meson contribution in
the amplitude
$A^{(e^+e^-\to\gamma\pi\pi)}_{\mu\alpha}(s_V\,,\,s_S\,,\,0)$  at
all (please replace $s_V$ in Eq. (\ref{phi}) by $s_S$ and then put
it into Eq. (\ref{four})). That is why, the author of Ref.
\cite{anisovich} adds the two new backgrounds, the two last terms
in Eqs. (\ref{four}) and (\ref{five}).  Actually, the  $B_0\left
(s_V\,,\,s_S\right )$ background is not required to avoid this
catastrophe, as is  evident from the further consideration.

The author of Ref. \cite{anisovich} thinks that the
$e^+e^-\to\phi\to\gamma\pi\pi$ amplitude is
\begin{eqnarray}
\label{vvae+e-phi} &&
A^{(e^+e^-\to\phi\to\gamma\pi\pi)}_{\mu\alpha}(s_V\,,\,s_S\,,\,0)=\left
(g_{\mu\alpha} - \frac{2q_\mu P_{V\alpha}}{s_V-s_S}\right )
\frac{G_{e^+e^-\to\phi}}{s_V-m_\phi^2 +
\imath\sqrt{s_V}\Gamma_\phi (s_V)}\nonumber\\[12pt]
    &&\times\Biggl [\frac{A_{\phi\to\gamma f_0}\left
(m_\phi^2\,,\,m_{f_0}^2\,,0\right )\,g_{f_0\to\pi\pi} }{s_S-M_0^2
+ \imath g^2_\pi\rho_{\pi\pi}(s_S)+ \imath g^2_K\rho_{K\bar
K}(s_S)}+ B_\phi\left (m_\phi^2\,,\,s_S\,,\,0\right )\Biggr ]
\end{eqnarray}
in accordance with Eq. (\ref{four}).

 But it is an illusion. To clear up the physical content of Eq.
 (\ref{four}) one should resolve the constraint (\ref{five}). Let
 us find $B_{f_0}\left(s_V\,,\,m_{f_0}^2\,,\,0\right)$ from  Eq.
 (\ref{five})
\begin{eqnarray}
\label{bf0}
&&B_{f_0}\left(s_V\,,\,m_{f_0}^2\,,\,0\right)\,g_{f_0\to\pi\pi}=-\frac{G_{e^+e^-\to\phi}\,A_{\phi\to\gamma
f_0}\left (m_\phi^2\,,\,m_{f_0}^2\,,0\right )
\,g_{f_0\to\pi\pi}}{s_V-M_\phi^2 + \imath\sqrt{s_V}\Gamma_\phi
(s_V)}\nonumber\\[9pt]
 && - \Biggl [s_V-M_0^2 +
\imath g^2_\pi\rho_{\pi\pi}(s_V)+ \imath g^2_K\rho_{K\bar
K}(s_V)\Biggr ]\Biggl [\frac{G_{e^+e^-\to\phi}\,B_\phi\left
(m_\phi^2\,,\,s_V\,,\,0\right )}{s_V-M_\phi^2 +
\imath\sqrt{s_V}\Gamma_\phi (s_V)}\nonumber\\
    &&+ B_0\left (s_V\,,\,s_V\right)\Biggr ]
\end{eqnarray}
 and  put it into
Eq. (\ref{four})
\begin{eqnarray}
\label{correct} &&
A^{(e^+e^-\to\gamma\pi\pi)}_{\mu\alpha}(s_V\,,\,s_S\,,\,0)=\left
(g_{\mu\alpha} - \frac{2q_\mu P_{V\alpha}}{s_V-s_S}\right
)\Biggl\{\frac{G_{e^+e^-\to\phi}}{s_V-M_\phi^2 +
\imath\sqrt{s_V}\Gamma_\phi (s_V)}\nonumber\\[9pt]
    && \times\Biggl [\frac{\left (s_S-s_V\right )B_\phi\left
(m_\phi^2\,,\,s_V\,,\,0\right )}{s_S-M_0^2+ \imath
g^2_\pi\rho_{\pi\pi}(s_S)+ \imath g^2_K\rho_{K\bar K}(s_S)
 } + B_\phi\left
(m_\phi^2\,,\,s_S\,,\,0\right )- B_\phi\left
(m_\phi^2\,,\,s_V\,,\,0\right ) \Biggr]\nonumber\\[9pt] && +
\frac{\left (s_S-s_V\right )B_0\left (s_V\,,\,s_V\,,\,0\right
)}{s_S-M_0^2+ \imath g^2_\pi\rho_{\pi\pi}(s_S)+ \imath
g^2_K\rho_{K\bar K}(s_S)
 } + B_0\left (s_V\,,\,s_S\,,\,0\right
)- B_0\left (s_V\,,\,s_V\,,\,0\right )\Biggr\}.
\end{eqnarray}
So, the amplitude  the $e^+e^-\to\phi\to\gamma\pi\pi$ amplitude is
\begin{eqnarray}
\label{nnae+e-phi} &&
A^{(e^+e^-\to\phi\to\gamma\pi\pi)}_{\mu\alpha}(s_V\,,\,s_S\,,\,0)=\left
(g_{\mu\alpha} - \frac{2q_\mu P_{V\alpha}}{s_V-s_S}\right
)\frac{G_{e^+e^-\to\phi}}{s_V-M_\phi^2 +
\imath\sqrt{s_V}\Gamma_\phi (s_V)}\nonumber\\[9pt]
    && \times\Biggl [\frac{\left (s_S-s_V\right )B_\phi\left
(m_\phi^2\,,\,s_V\,,\,0\right )}{s_S-M_0^2+ \imath
g^2_\pi\rho_{\pi\pi}(s_S)+ \imath g^2_K\rho_{K\bar K}(s_S)
 } + B_\phi\left
(m_\phi^2\,,\,s_S\,,\,0\right )- B_\phi\left
(m_\phi^2\,,\,s_V\,,\,0\right ) \Biggr]
\end{eqnarray}
and not (\ref{vvae+e-phi}).

As is seen from Eq. (\ref{nnae+e-phi}) the $\phi\gamma f_0$ vertex
has a classic form
\begin{eqnarray}
\label{f0classic} && V_{\phi\gamma f_0}=
e^\mu(\phi)e^\alpha(\gamma)\left (g_{\mu\alpha} - \frac{2q_\mu
P_{V\alpha}}{s_V-s_S}\right )\left (s_S-s_V\right )B_\phi\left
(m_\phi^2\,,\,s_V\,,\,0\right )\nonumber\\[9pt] &&= - 2P_V^\nu
e^\mu(\phi)F_{\nu\mu}B_\phi\left (m_\phi^2\,,\,s_V\,,\,0\right )
\end{eqnarray}
where $F_{\nu\mu}=q_\mu e_\nu (\gamma)-q_\nu e_\mu (\gamma)$,
$e(\phi)$ and $e(\gamma)$ are the polarization four-vectors of the
$\phi$ meson and the $\gamma$ quantum, respectively, $B_\phi\left
(m_\phi^2\,,\,s_V\,,\,0\right )$ is an invariant vertex  free from
kinematical singularities. All of the preceding concerns the
backgrounds
\begin{eqnarray}
\label{bphiclassic} && A_{\phi\gamma\pi\pi}=
e^\mu(\phi)e^\alpha(\gamma)\left (g_{\mu\alpha} - \frac{2q_\mu
P_{V\alpha}}{s_V-s_S}\right )\Biggl [ B_\phi\left
(m_\phi^2\,,\,s_S\,,\,0\right )- B_\phi\left
(m_\phi^2\,,\,s_V\,,\,0\right )\Biggr ]\nonumber\\[9pt]
&&=2P_V^\nu e^\mu(\phi)F_{\nu\mu}\,\frac{B_\phi\left
(m_\phi^2\,,\,s_S\,,\,0\right )- B_\phi\left
(m_\phi^2\,,\,s_V\,,\,0\right )]} {s_V-s_S}
\end{eqnarray}
and
\begin{eqnarray}
\label{b0classic} && \tilde A^{(e^+e^-\to\gamma\pi\pi)}_\mu
(s_V\,,\,s_S\,,\,0)= e^\alpha (\gamma)\left (g_{\mu\alpha} -
\frac{2q_\mu P_{V\alpha}}{s_V-s_S}\right )\nonumber\\[9pt]
&&\times\Biggl
 [\frac{\left (s_S-s_V\right )B_0\left (s_V\,,\,s_V\,,\,0\right
)}{s_S-M_0^2+ \imath g^2_\pi\rho_{\pi\pi}(s_S)+ \imath
g^2_K\rho_{K\bar K}(s_S)
 } + B_0\left (s_V\,,\,s_S\,,\,0\right )- B_0\left (s_V\,,\,s_V\,,\,0\right
)\Biggr ]\nonumber\\[9pt] && = - 2P_V^\nu
F_{\mu\nu}\Biggl[\frac{B_0\left (s_V\,,\,s_V\,,\,0\right
)}{s_S-M_0^2+ \imath g^2_\pi\rho_{\pi\pi}(s_S)+ \imath
g^2_K\rho_{K\bar K}(s_S)
 }\nonumber\\[9pt]
 && + \frac{B_0\left
(s_V\,,\,s_V\,,\,0\right )- B_0\left (s_V\,,\,s_S\,,\,0\right
)}{s_V-s_S}\Biggr ].
\end{eqnarray}

That the $\phi\gamma f_0$ vertex and the $A_{\phi\gamma\pi\pi}$
background follow the same function $B_\phi\left
(m_\phi^2\,,\,x\,,\,0\right )$, see Eqs.  (\ref{f0classic}) and
(\ref{bphiclassic}), is an artifact  rather than  an achievement
of the theory under discussion. It hardly needs proposing the
experimental investigation of this phenomenon \cite{sigma}.
Really, one can add a regular term to the right side of Eq.
(\ref{four}), for example,
\begin{eqnarray}
\label{forexample} && \left (g_{\mu\alpha} - \frac{2q_\mu
P_{V\alpha}}{s_V-s_S}\right )\frac{G_{e^+e^-\to\phi}}{s_V-M_\phi^2
+ \imath\sqrt{s_V}\Gamma_\phi (s_V)}\Biggl [A_\phi\left
(m_\phi^2\,,\,s_V\,,\,0\right )- A_\phi\left (m_\phi^2\,,\,s_S
\,,\,0\right ) \Biggr]
\end{eqnarray}
which does not contribute to Eqs. (\ref{five}) and (\ref{bf0}) but
change Eqs. (\ref{correct}), (\ref{nnae+e-phi}), and
(\ref{bphiclassic})
\begin{eqnarray}
\label{nnae+e-phia} &&
A^{(e^+e^-\to\phi\to\gamma\pi\pi)}_{\mu\alpha}\left(s_V\,,\,s_S\,,\,0\right)\to
\left (g_{\mu\alpha} - \frac{2q_\mu P_{V\alpha}}{s_V-s_S}\right
)\frac{G_{e^+e^-\to\phi}}{s_V-M_\phi^2 +
\imath\sqrt{s_V}\Gamma_\phi (s_V)}\nonumber\\[9pt]
    && \times\Biggl \{\frac{\left (s_S-s_V\right )B_\phi\left
(m_\phi^2\,,\,s_V\,,\,0\right )}{s_S-M_0^2+ \imath
g^2_\pi\rho_{\pi\pi}(s_S)+ \imath g^2_K\rho_{K\bar K}(s_S)
 }\nonumber\\[9pt]
 && + \Biggl [B_\phi\left
(m_\phi^2\,,\,s_S\,,\,0\right )-A_\phi\left
(m_\phi^2\,,\,s_S\,,\,0\right )\Biggr ] - \Biggl [B_\phi\left
(m_\phi^2\,,\,s_V\,,\,0\right )- A_\phi\left
(m_\phi^2\,,\,s_V\,,\,0\right )\Biggl ] \Biggr \}
\end{eqnarray}
and
\begin{eqnarray}
\label{bphiclassica} && A_{\phi\gamma\pi\pi}\to
e^\mu(\phi)e^\alpha(\gamma)\left (g_{\mu\alpha} - \frac{2q_\mu
P_{V\alpha}}{s_V-s_S}\right )\nonumber\\[9pt] &&\times\Biggl \{
\Biggl [B_\phi\left (m_\phi^2\,,\,s_S\,,\,0\right )- A_\phi\left
(m_\phi^2\,,\,s_S\,,\,0\right )\Biggr ]- \Biggl [B_\phi\left
(m_\phi^2\,,\,s_V\,,\,0\right )- A_\phi\left
(m_\phi^2\,,\,s_V\,,\,0\right )\Biggr ]\Biggr \}\nonumber\\[9pt]
&&=2P_V^\nu e^\mu(\phi)F_{\nu\mu}\nonumber\\[9pt] &&
\times\frac{\biggl [B_\phi\left (m_\phi^2\,,\,s_S\,,\,0\right )-
A_\phi\left (m_\phi^2\,,\,s_S\,,\,0\right )\biggr ]- \biggl
[B_\phi\left (m_\phi^2\,,\,s_V\,,\,0\right )- A_\phi\left
(m_\phi^2\,,\,s_V\,,\,0\right )\biggr ]} {s_V-s_S}.
\end{eqnarray}

So, if $B_\phi\left (m_\phi^2\,,\,x\,,\,0\right )= A_\phi\left
(m_\phi^2\,,\,x\,,\,0\right )$ the background (\ref{bphiclassica})
is removed from Eq. (\ref{nnae+e-phia}) at all. The background
(\ref{b0classic}) can be removed also by  the proper regular term
in Eq. (\ref{four}). Consequently, the inclusion of the
backgrounds is not a kategorischer Imperativ, but a
phenomenological treatment.

Nevertheless, let us compare the $\phi\to\gamma\pi\pi$ decay
amplitude from Eq. (\ref{nnae+e-phi}) at $s_V=m_\phi^2$
\begin{eqnarray}
\label{nnaphigammapipi} && A^{(\phi\to\gamma\pi\pi)}\left
(m_\phi^2\,,\,s_S\,,\,0\right)=\frac{\left (s_S-m_\phi^2\right
)B_\phi\left (m_\phi^2\,,\,m_\phi^2\,,\,0\right )}{s_S-M_0^2+
\imath g^2_\pi\rho_{\pi\pi}(s_S)+ \imath g^2_K\rho_{K\bar K}(s_S)
 }\nonumber\\[9pt]
 && + B_\phi\left
(m_\phi^2\,,\,s_S\,,\,0\right )- B_\phi\left
(m_\phi^2\,,\,m_\phi^2\,,\,0\right )
\end{eqnarray}
with the one from Eq. (\ref{vvae+e-phi})
\begin{eqnarray}
\label{vvaphigammapipi} && A^{(\phi\to\gamma\pi\pi)}_{VVA}\left
(m_\phi^2\,,\,s_S\,,\,0\right )=\frac{A_{\phi\to\gamma f_0}\left
(m_\phi^2\,,\,m_{f_0}^2\,,0\right )\,g_{f_0\to\pi\pi} }{s_S-M_0^2
+ \imath g^2_\pi\rho_{\pi\pi}(s_S)+ \imath g^2_K\rho_{K\bar
K}(s_S)}\nonumber\\[9pt] &&+ B_\phi\left
(m_\phi^2\,,\,s_S\,,\,0\right )\,,
\end{eqnarray}
which the author of Ref. \cite{anisovich} considers together with
the extra constraint
\begin{eqnarray}
\label{constraint}
 \frac{A_{\phi\to\gamma f_0}\left
(m_\phi^2\,,\,m_{f_0}^2\,,0\right )\,g_{f_0\to\pi\pi}
}{m_\phi^2-M_0^2 + \imath g^2_\pi\rho_{\pi\pi}(m_\phi^2)+ \imath
g^2_K\rho_{K\bar K}(m_\phi^2)}+ B_\phi\left
(m_\phi^2\,,\,m_\phi^2\,,\,0\right )=0
\end{eqnarray}
to provide the $O\left (m_\phi^2-s_S\right )$ behavior of
$A^{(\phi\to\gamma\pi\pi)}_{VVA}\left
(m_\phi^2\,,\,s_S\,,\,0\right )$ at $ \left (m_\phi^2-s_S\right
)\to 0$ (for soft photons).

It follows from Eqs. (\ref{vvaphigammapipi}) and
(\ref{constraint}) that
\begin{eqnarray}
\label{physvvaphigammapipi} &&
A^{(\phi\to\gamma\pi\pi)}_{VVA}\left
(m_\phi^2\,,\,s_S\,,\,0\right)\nonumber\\[9pt] &&=\frac{\biggl
[s_S-m_\phi^2 + \imath g^2_K\rho_{K\bar K}(s_S) + \imath
g^2_\pi\rho_{\pi\pi}(s_S)-\imath g^2_\pi\rho_{\pi\pi}(m_\phi^2)-
\imath g^2_K\rho_{K\bar K}(m_\phi^2)\biggr ]B_\phi\left
(m_\phi^2\,,\,m_\phi^2\,,\,0\right )}{s_S-M_0^2+ \imath
g^2_\pi\rho_{\pi\pi}(s_S)+ \imath g^2_K\rho_{K\bar K}(s_S)
 }\nonumber\\[9pt]
 && + B_\phi\left
(m_\phi^2\,,\,s_S\,,\,0\right )- B_\phi\left
(m_\phi^2\,,\,m_\phi^2\,,\,0\right )\,,
\end{eqnarray}
where \cite{anisovich05}
\begin{eqnarray}
\label{anisnotification} &&
\rho_{\pi\pi}=\frac{1}{M_0}\sqrt{M_{\pi\pi}^2-4m_\pi^2}\,,\ \
\rho_{K\bar K}=\frac{1}{M_0}\sqrt{M_{\pi\pi}^2-4m_K^2}\,\ \
\mbox{when}\ \ M_{\pi\pi}>2m_K\,,\nonumber\\[6pt]
 && \ \rho_{K\bar
K}=\imath \frac{1}{M_0}\sqrt{4m_K^2-M_{\pi\pi}^2}\,\ \
\mbox{when}\ \ M_{\pi\pi}<2m_K\,, \nonumber\\[6pt]
 &&
g_\pi^2=0.12\,\mbox{GeV}^2,\ \ \ g_K^2=0.27\,\mbox{GeV}^2,\ \ \
M_0=0.975\,\mbox{GeV}\,,
\end{eqnarray}
in addition,  \cite{anisovich05}
\begin{equation}
\label{bphi} B_\phi\left (m_\phi^2\,,\,s_S\,,\,0\right
)=B_\phi\left (m_\phi^2\,,\,m_\phi^2\,,\,0\right )\biggl [1+a\left
(s_S-m_\phi^2 \right )\biggr ]\exp \Biggl [-\frac{m_\phi^2-s_S
}{\mu^2}\Biggr ]
\end{equation}
where $a=-5\,\mbox{GeV}^{-2}$ and $\mu=0.388\,\mbox{GeV}$.

The calculation gives
\begin{equation}
\label{branchings}
 BR\left (\phi\to\gamma\pi\pi\,\mbox{by Eq.
(14)}\right )/BR\left (\phi\to\gamma\pi\pi\,\mbox{by Eq.
(17)}\right )\approx 0.2
\end{equation}
 and for
 \begin{eqnarray}
 \label{spectra}
&&\frac{dBR\left (\phi\to\gamma\pi\pi\,\mbox{by Eq.
(14)}\,,\,\sqrt{s_S}\right )}{d\sqrt{s_S}}\Biggm /\frac{dBR\left
(\phi\to\gamma\pi\pi\,\mbox{by Eq. (17)}\,,\,\sqrt{s_S}\right
)}{d\sqrt{s_S}}\nonumber\\[9pt] &&=R\left (\sqrt{s_S}\, \right ):
 \end{eqnarray}
 \begin{eqnarray}
 \label{spectraratios}
 && R(800\,\mbox{MeV})=0.094\,,\ R(850\,\mbox{MeV})=0.131\,,\
 R(900\,\mbox{MeV})=0.191\,,\nonumber\\
 && R(950\,\mbox{MeV})=0.249\,,\ R(970\,\mbox{MeV})=0.266\,,\
 R(990\,\mbox{MeV})=0.323\,.
 \end{eqnarray}
So, both the branching ratios and the spectra are essentially
different in case of Eqs. (\ref{nnaphigammapipi}) and
(\ref{physvvaphigammapipi}).

{\bf Last comment.} The author of Ref. \cite{anisovich} thinks,
see Eq. (12) in Ref. \cite{anisovich}, that the phase of the
$\phi\to\gamma\pi\pi$ amplitude equals to the $\pi\pi$ scattering
phase shift $\delta^0_0$. But it is a misunderstanding. The Watson
theorem is not correct in our case because  there are other open
intermediate channels, even though $s_S<(2m_K)^2$.  The $\phi\to
K^+K^-\to\gamma\pi\pi$ channel is most essential  from them, see
details in Ref. \cite{achasov03}.

{\bf Summary.} The $\phi\to\gamma\pi\pi$ amplitude
(\ref{nnaphigammapipi}), which we found from Eqs. (\ref{four}) and
(\ref{five}) suggested in Ref. \cite{anisovich}, differs
essentially from the $\phi\to\gamma\pi\pi$ amplitude
(\ref{physvvaphigammapipi}),  which is studied  in Refs.
\cite{anisovich,anisovich05}. There are no reasons which prevent
us from adding a regular term  to the right side of Eq.
(\ref{four}) without changing  Eq. (\ref{five}) but with changing
backgrounds, that is,  backgrounds are included phenomenologically
( by hands) and not dictated by resonance properties. The idea of
the author Ref. \cite{anisovich} about the phase of the
$\phi\to\gamma\pi\pi$ decay amplitude is misleading.


\begin{thebibliography}{99}
\bibitem{sigma}
The author of  Ref. \cite{anisovich} would like to connect the
$B_\phi\left (m_\phi^2\,,\,s_S\,,\,0\right )$ background with
isoscalar scalar states in the $\pi\pi$ channel    other than
$f_0(980)$ " (for example, such as $\sigma$)", see Section 1.2,
page 4 of Ref. \cite{anisovich}. But in that case the author of
Ref. \cite{anisovich} should predict the production of the dual
vector states in the $e^+e^-$ collision, for example, such as
$V(m_\sigma)$, that would be described by $B_\phi\left
(m_\phi^2\,,\,s_V\,,\,0\right )$ in the $e^+e^-\to\gamma\pi\pi$
amplitude (\ref{correct}).
\end{thebibliography}
\end{document}